\documentclass[%
reprint,
superscriptaddress,
 amsmath,
 amssymb,
pra,
floatfix,
]{revtex4-2}

\usepackage{suli}
\usepackage{jabbrv} 

\usepackage{lineno}

\usepackage[absolute]{textpos}
\usepackage{xcolor}

\definecolor{darkblue}{rgb}{0,0,0.5}
\hypersetup{
colorlinks=true,
linkcolor=black,
filecolor=blue,
citecolor=darkblue,
urlcolor=black,
}

\graphicspath{{Images/}}

\begin{document}

\title{Optimal resource allocation for flexible-grid entanglement distribution networks}

\author{Jude Alnas}
    \affiliation{Department of Electrical and Computer Engineering, University of Alabama, 245 7th Avenue, Tuscaloosa, Alabama 35406, USA}
    \affiliation{Quantum Information Science Section, Oak Ridge National Laboratory, Oak Ridge, Tennessee 37831, USA}
\author{Muneer Alshowkan}
    \affiliation{Quantum Information Science Section, Oak Ridge National Laboratory, Oak Ridge, Tennessee 37831, USA}
\author{Nageswara~S.~V. Rao}
    \affiliation{Advanced Computing Methods Section, Oak Ridge National Laboratory, Oak Ridge, Tennessee 37831, USA}
\author{Nicholas~A. Peters} 
    \affiliation{Quantum Information Science Section, Oak Ridge National Laboratory, Oak Ridge, Tennessee 37831, USA}
\author{Joseph~M. Lukens}
    \email{lukensjm@ornl.gov} 
    \affiliation{Quantum Information Science Section, Oak Ridge National Laboratory, Oak Ridge, Tennessee 37831, USA}

\date{\today}

\begin{abstract}
We use a genetic algorithm (GA) as a design aid for determining the optimal provisioning of entangled photon spectrum in flex-grid quantum networks with arbitrary numbers of channels and users. After introducing a general model for entanglement distribution based on  frequency-polarization hyperentangled biphotons, we derive upper bounds on fidelity and entangled bit rate for networks comprising one-to-one user connections. Simple conditions based on user detector quality and link efficiencies are found that determine whether entanglement is possible. We successfully apply a GA to find optimal resource allocations in four different representative network scenarios and validate features of our model experimentally in a quantum local area network in deployed fiber. Our results show promise for the rapid design of large-scale entanglement distribution networks.
\end{abstract}


\maketitle

\section{Introduction}\label{sec:intro}
The phenomenon of quantum entanglement has resulted in a plethora of novel quantum technologies. Examples in quantum information science (QIS) include quantum-secured communications~\cite{Bennett1992,Ekert1991,Tittel2000}, blind or distributed quantum computing~\cite{Stefanie2012,Fitzsimons2017}, and the emerging quantum internet~\cite{Kimble2008, Wehner2018}. Entanglement also promises new techniques beyond computing and communications, such as improved interferometry for radio astronomy~\cite{Gottesman2012,Khabiboulline2019} and entanglement-enhanced clock synchronization~\cite{Jozsa2000,Komar2014} for sensing and navigation~\cite{Giovannetti2001}. Realizing these technologies will require robust quantum networks, the infrastructure with which quantum information can be transmitted and received. Such a network must also be capable of reliably distributing entangled qubits to network users on demand. Early examples of entanglement distribution for quantum communications can be found in quantum key distribution (QKD) experiments~\cite{Tittel2000,Ursin2007,Dynes2009,Ribordy2000} that, however, focused on only two directly connected users. 

The development of spontaneous parametric down-conversion (SPDC) techniques has facilitated the generation of telecommunication wavelength ($\sim$1550~nm) photons entangled in both the polarization and frequency degrees of freedom~\cite{Jiang2006, Kaiser2012, Vergyris2017, Yamazaki2021, Ponce2022}. Using frequency entanglement, optical techniques from classical networks such as wavelength division multiplexing (WDM) have been leveraged to perform entanglement distribution, forming rudimentary quantum local area networks (QLANs). Much like conventional routing protocols such as the Transmission Control Protocol/Internet Protocol (TCP/IP), the working principles of these QLANs can be described in layers of abstraction \cite{Wengerowsky2018, Alshowkan2021, Chung2021, Pompili2021} as in \cref{fig:qlan_layers}. Under our proposed definitions~\cite{Alshowkan2021}, the physical layer consists of a hyperentangled light source realized through a pump laser and a nonlinear optical crystal~\cite{PhysRevLett.95.260501} or a waveguide capable of SPDC. This spectrum then enters a wavelength demultiplexer (demux). The link layer is the next abstraction layer and is implemented entirely within the demux, which is configured to partition a hyperentangled spectrum into pairs of frequency-entangled bands. In the subsequent allocation process, users are selectively entangled by multiplexing energy-correlated channel pairs onto their respective physical ports. Users receiving correlated channels share entanglement via the polarization degree of freedom, thus creating a logical network of entanglement that constitutes the transport layer.

Early instances of this approach used fixed-grid technology such as dense wavelength division multiplexing (DWDM) filters to implement the link layer~\cite{Marcikic2003,Lim2008,Zhou2013,Wengerowsky2018,Wengerowsky2019,Joshi2020}. In such a setup, the hyperentangled spectrum is partitioned into a fixed number of fixed-width frequency channels. Likewise, the routing of these channels is essentially fixed, modifiable only by physically disconnecting and reconnecting the DWDM filters or by using supplementary spatial optical switches~\cite{Peters2009, Herbauts2013}. 
Increasing network capacity requires extra frequency channels, which in turn require additional DWDM and switching components. For example, in \cite{Wengerowsky2019}, a fully connected $N$-user network would require $4N-1$ additional filters to support one more user. Because these channels are fixed in bandwidth, one also faces the potential problem of inefficient spectrum utilization, whereby users receive either more or less bandwidth than required for their desired application. 


Flexible wavelength-selective switches (WSSs) based on liquid crystal on silicon technology~\cite{Roelens2008} avoid many of the drawbacks of fixed-grid networks. A WSS provides active wavelength multiplexing and routing in a single device while also allowing for optimal bandwidth utilization. In the context of entanglement distribution, such a device can be used to partition a spectrum of entangled light into an arbitrary number of frequency channels of arbitrary widths (within device limitations), obviating the need for numerous passive components. Channels are then allocated to a physical path connected to a user or a second WSS. Routing is done electronically within the device, allowing rapid reconfiguration of entanglement networks without discrete switching components. As a result, flex-grid entanglement distribution networks are more compact, efficient, and adaptable in comparison to their fixed-grid counterparts. This technology has already found use in recent demonstrations of entanglement distribution in QLANs~\cite{Lingaraju2021,Appas2021,Alshowkan2021,Alshowkan2021b}. 

The dynamic partitioning and allocation facilitated by WSSs also allow for entanglement network optimization. For example, allocations can be adjusted to compensate for losses from low-quality detectors~\cite{Lingaraju2021} or long-distance transmission~\cite{Appas2021}. However, problems arise in finding the proper frequency allocations for optimal entanglement distribution. A trial-and-error approach was used in \cite{Lingaraju2021} to find allocations of multiple fixed-width channels to equalize coincidence counts in a fully connected four-user network. In \cite{Appas2021}, an undescribed algorithm and model was used to find the proper channel widths in a fixed allocation to equalize coincidence counts in a five-user network where one user suffered losses due to long-distance transmission. While trial-and-error or exhaustive search methods may be viable for small networks, they are intractable at large scales. Hence, there is a need for tools with which entanglement resource allocations can be easily found to meet global network objectives.

In this paper, we propose a model of an entanglement network for use in tackling the entangled flux allocation (EFA) problem. In \cref{sec:model}, we introduce the many-to-many entanglement network model in terms of biphoton fluxes, specialize to the case of one-to-one networks, and derive upper bounds and conditions on entanglement fidelity and entangled bit rate (EBR). In \cref{sec:network_opt}, we introduce the EFA optimization problem and demonstrate allocation optimization through the use of a genetic algorithm (GA) \cite{Holland1992} for a number of illustrative scenarios. We then follow in \cref{sec:exp} with an experimental test of our basic link model in a deployed QLAN on the Oak Ridge National Laboratory campus. Finally, implications and future avenues for research are discussed in \cref{sec:conclusions}.

\section{Entanglement network model}\label{sec:model}
\subsection{Fundamental model}\label{sec:fund_model}
\begin{figure*}[t]
    \centering
    \includegraphics[width=\linewidth]{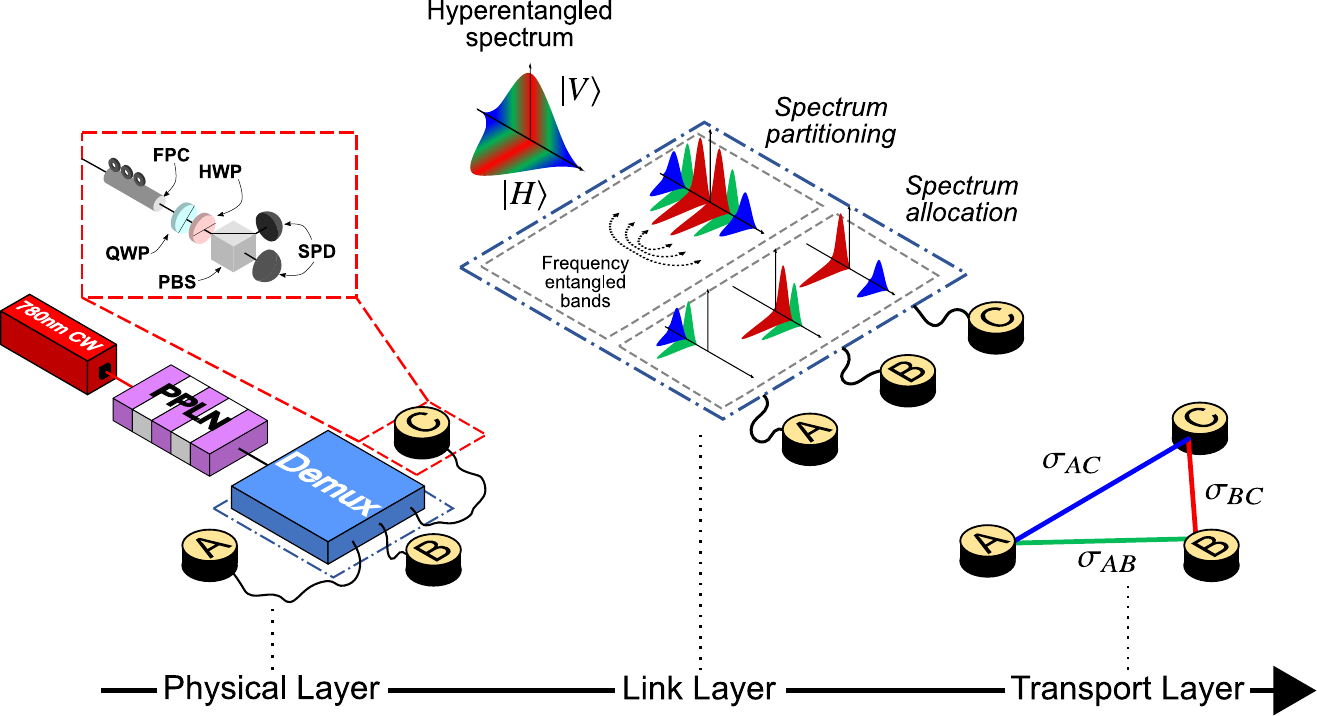}
    \caption{Example entanglement distribution network of $N=3$ users and $L=3$ links. Each user is equipped with a polarization analyzer (red inset). In the physical layer, a hyperentangled spectrum is generated using a continuous-wave (CW) 780~nm pump laser and a periodically poled lithium niobate (PPLN) waveguide. The process of spectrum partitioning and allocation occurs within the wavelength-selective demultiplexing device (demux) and constitutes the link layer. Here, the hyperentangled spectrum is partitioned into $K=3$ pairs of frequency-correlated bands and allocated to users, forming the logical network of entanglement links that constitutes the transport layer. 
    FPC: fiber polarization controller. QWP: quarter-wave plate. HWP: half-wave plate. PBS: polarizing beam-splitter. SPD: single-photon detector.}
    \label{fig:qlan_layers}
\end{figure*}

A polarization entanglement network of $N\in\mathbb{N}^+=\{1,2,...\}$ users and $L\in\mathbb{N}^+$ entanglement links can be represented by an undirected graph of $N$ nodes and $L$ edges as in the transport layer of \cref{fig:qlan_layers}. Each edge represents shared entanglement over link $l$, with density matrix $\sigma_l$.
%
In our model, link state $\sigma_l$ is calculated from quantities measurable in a polarized photon counting experiment between users $m_l$ and $n_l$, namely, the rate of accidental coincidences $A_l$ and the rate of correlated coincidences $C_l$. In this experiment, it is assumed that each user has two thresholding photon detectors of identical system efficiency $\eta$ and dark count rate $d$ configured to count photons of orthogonal polarizations in some basis (e.g., rectilinear, diagonal, or circular). Note that $\eta$ includes all losses due to optical components and $d$ includes all ambient light apart from the allocated biphoton channels.  For example, user $m_l$ on link $l$ has two detectors of efficiency $\eta_{m_l}$ and dark count rate $d_{m_l}$ (variation \emph{between} users is allowed, i.e., $\eta_{m_l}\neq\eta_{n_l}$ and $d_{m_l}\neq d_{n_l}$). 
With these definitions, the twofold accidental coincidence rate $A_l$, summed over all four detector pair combinations, can be written as~\cite{Eckart1938,Pearson2010}
%
\begin{equation}\label{eqn:accidentals}
    A_l=4\tau\left(\frac{\eta_{m_l}}{2}\sum_{l'\in\mathcal{L}_{m_l}}\bar{\mu}_{l'}+d_{m_l}\right) \left(\frac{\eta_{n_l}}{2}\sum_{l'\in\mathcal{L}_{n_l}}\bar{\mu}_{l'}+d_{n_l}\right).
\end{equation}
The correlated coincidence rate---i.e., from detection events corresponding to photons from the same entangled pair---is then
%
\begin{equation}\label{eqn:true_coincidences}
    C_l=\eta_{m_l}\eta_{n_l}\bar{\mu}_l,
\end{equation}
equal to the total coincidences that would be obtained in the absence of background noise and multi-pair emission. In \cref{eqn:accidentals,eqn:true_coincidences}, $\tau$ is the coincidence window (assumed wide enough to integrate over the entire correlated coincidence peak), and $\mathcal{L}_{m_l}$ is the set of links connected to user $m_l$. Flux $\bar{\mu}_l$ is the total biphoton flux over all channels allocated to users $m_l$ and $n_l$ on link $l$, defined at the source (i.e., prior to system loss).

The two-qubit polarization state of the link, $\sigma_l$, in the coincidence basis for the pair of users, 
is then expressed as the following density matrix:
%
\begin{equation}\label{eqn:full_model}
    \sigma_l=\lambda_l
    \sum_{k\in\mathcal{B}_l}\frac{\mu_k}{\bar{\mu}_l}U_k\rho_kU^\dagger_k+
    (1-\lambda_l)\frac{I_4}{4},
\end{equation}
where $\lambda_l$ is the visibility given by
%
\begin{equation}\label{eqn:lambda}
    \lambda_l = \frac{C_l}{A_l+C_l}.
\end{equation}
In this model, the spectrum of entangled light is partitioned into $2K$ entangled frequency bands paired into $K$ frequency channels (illustrated in \cref{fig:qlan_layers}); thus, $K\geq L$ is a prerequisite for allocating flux to all desired links. An entangled photon source emits photon pairs in channel $k\in\{1,2,\dots,K\}$ at a flux of $\mu_k$, described by the density matrix $\rho_k$; unitary matrix $U_k$ encapsulates any propagation effects (e.g., birefringence) experienced by the photons. It is important to emphasize that $\rho_k$ models the state of exactly two photons (originating from the same pump photon); nonidealites from multi-pair effects and background counts appear through \cref{eqn:accidentals,eqn:true_coincidences}. In constructing the complete (noisy) state $\sigma_l$, the $\rho_k$ are weighted by the ratio of channel flux $\mu_k$ to total link flux $\bar{\mu}_l=\sum_{k\in\mathcal{B}_l}\mu_k$, where $\mathcal{B}_l\subseteq\{1,2,\dots,K\}$ is the set of channels allocated to link $l$. We assume all noise in the network is uniform background modeled by the maximally mixed two-qubit state $I_4/4$ where $I_4$ is the $4\times 4$ identity matrix. When correlated coincidences dominate, $\lambda_l\to1$ and $\sigma_l$ approaches a convex sum of entangled states. Conversely, when accidentals dominate, $\lambda_l\to0$ and $\sigma_l$ approaches the maximally mixed state.

The metrics chosen to quantify network performance are the link state fidelities $F_l$ with respect to some target state $\ket{\Psi_l}$ and the entangled bit rates (EBRs) $R_l$, defined as:
%
\begin{equation}\label{eqn:fidelity_braket}
    F_l=\bra{\Psi_l}\sigma_l\ket{\Psi_l},
\end{equation}
\begin{equation}\label{eqn:ebr_logneg}
    R_{l}=(A_l+C_l)\log_2(||\sigma_l^{T_A}||_1),
\end{equation}
where the logarithmic term is the logarithmic negativity [Eq.~(2) in~\cite{Vidal2002}], $\sigma_l^{T_A}$ is the partial transpose of the link state with respect to one photon of the entangled pair, and $||\cdot||_1$ is the trace norm.
%
In \cref{eqn:ebr_logneg}, $R_l\in\mathbb{R}_{\geq0}$ is an upper bound on the rate of distillable entanglement received by the users of link $l$~\cite{Vidal2002}. Intuitively, $F_l$ assesses the quality of two coincident photon detections, whereas $R_l$ incorporates both quality and quantity (rate) into a single number.

Before proceeding further, it is useful to note the regimes under which the model expressed in Eqs.~(\ref{eqn:accidentals}--\ref{eqn:full_model}) remains valid. First, the probability of a single detector click within the coincidence window $\tau$ must be much less than unity: e.g., for user $m_l$, $\tau(\frac{\eta_{m_l}}{2}\sum_{l\in\mathcal{L}_{m_l}}\bar{\mu}_l+d_{m_l}) \ll 1$ (and analogously for all other users). This condition ensures that the rate of accidental coincidences follows the standard ``product-of-singles'' formula~\cite{Eckart1938, Pearson2010}. Second, the singles detection rates must remain constant regardless of polarization analyzer settings, implying that the traced out marginal quantum state for each user possesses an equal distribution of $H$ and $V$ components; this holds for all Bell and Werner states, for example. Finally, the correlated coincidence rate in \cref{eqn:true_coincidences} assumes photon pairs that are independent---i.e., distinguishable---of each other, a situation occurring whenever the pump laser in the SPDC process has much lower bandwidth than the narrowest frequency channels, which has been the case in all flex-grid entanglement experiments so far~\cite{Lingaraju2021,Alshowkan2021,Appas2021}.
Importantly, however, the general approach we introduce  below for optimal bandwidth allocation is not limited to the current expressions and assumptions, but can be adapted to any physical model as long as it connects fixed user parameters (e.g., $\eta$ and $d$) and allocation decisions (e.g., $\mathcal{B}_l$) to the quantities of interest (e.g., $F_l$ and $R_l$).

\subsection{Simplified one-link model}\label{sec:simple_model}
In order to highlight the important features of our method with minimal distractions, we apply three additional assumptions to the model described in \cref{sec:fund_model} for the network scenarios simulated below. 
%
\begin{assumption}\label{assump:121}
Users are entangled with only one other. \\ $|\mathcal{L}_{m_l}|=|\mathcal{L}_{n_l}|=1~\forall~l\in\{1,2,...,L\}$.
\end{assumption}
%
\begin{assumption}\label{assump:psi_k}
Channel states are identical. \\ $\rho_k=\ketbra{\psi}{\psi}~\forall~k\in\{1,2,...,K\}$.
\end{assumption}
%
\begin{assumption}\label{assump:unitaries}
Channel distortion effects are fully compensated. \\  $U_k=I_4~\forall~k\in\{1,2,...,K\}$.
\end{assumption}
\Cref{assump:121} restricts the logical network topology to one-to-one links, making it sufficient to analyze a single link in isolation to understand the basic behavior. Accordingly, in the initial analysis, we suppress the $l$ subscripts and treat a single link of flux $\bar{\mu}$ consisting of user 1 and user 2, characterized by efficiencies $\eta_1$, $\eta_2$ and dark count rates $d_1$, $d_2$. 
Although simpler than the fully connected paradigm, whereby each user is entangled with all others in a single physical configuration \cite{Wengerowsky2018, Joshi2020, Lingaraju2021, Appas2021}, the one-to-one restriction is more naturally suited to flex-grid capabilities in our view. For, as argued in \cite{Alshowkan2021}, the fact a WSS can be reconfigured on demand to realize entanglement between any two users obviates the need to permanently dedicate a wavelength channel to every pair, improving scalability and reducing crosstalk effects by removing the summations in \cref{eqn:accidentals}. The accidental rate $A$ can then be expressed as a polynomial in a single flux value $\bar{\mu}$. Specifically,
\begin{equation}\label{eqn:simple_accidentals}
    A=4\tau\left[\frac{\eta_1\eta_2}{4}\bar{\mu}^2+\left(\frac{\eta_1}{2}d_2+\frac{\eta_2}{2}d_1\right)\bar{\mu}+d_1d_2\right].
\end{equation}
\Cref{assump:psi_k,assump:unitaries} eliminate the summation in \cref{eqn:full_model}, leaving the Werner state
\begin{equation}\label{eqn:simple_model}
    \sigma=\lambda\ketbra{\psi}{\psi}+(1-\lambda)\frac{I_4}{4},
\end{equation}
where, for concreteness, we assume the specific Bell state $\ket{\psi}=\ket{\Psi^-}\propto\ket{HV}-\ket{VH}$.

Physically speaking, these assumptions imply uniform entanglement across all frequency channels of interest and perfect compensation of any birefringence effects in the optical channel. Depending on the nature of the source and transmission medium, these assumptions may be difficult to attain, although they certainly represent the desired situation in practice. For our purposes, they allow us to focus on fundamental effects due to probabilistic photon emission without complications from technical nonidealities, which can be incorporated as needed to reflect an actual experiment.

For states of the form of \cref{eqn:simple_model}, fidelity with respect to $\ket{\psi}$ is simply $F=(1+3\lambda)/4$.
%
%
By Eqs.~(\ref{eqn:true_coincidences},\ref{eqn:lambda},\ref{eqn:simple_accidentals}), fidelity can then be expressed as a function of link flux $\bar{\mu}$ 
\begin{equation}\label{eqn:simple_fidelity}
    F(\bar{\mu})=\frac{1}{4}\left\{1+\frac{3\bar{\mu}}{
    4\tau\left[\frac{\bar{\mu}^2}{4}+(\frac{d_1}{2\eta_1}+\frac{d_2}{2\eta_2}+\frac{1}{4\tau})\bar{\mu}+\frac{d_1d_2}{\eta_1\eta_2}\right]
    }\right\}.   
\end{equation}
A similar expression for EBR can be obtained. For states of the form of \cref{eqn:simple_model}, the logarithmic negativity becomes a simple function of $F$ and thus a function of $\bar{\mu}$~\cite{Vidal2002}:
%
\begin{widetext}

\begin{equation}\label{eqn:simple_ebr}
    \begin{split}
        R(\bar{\mu})=4\tau\left[\frac{\eta_1\eta_2}{4}\bar{\mu}^2+\left(\frac{\eta_1}{2}d_2+\frac{\eta_2}{2}d_1+\frac{\eta_1\eta_2}{4\tau}\right)\bar{\mu}+d_1d_2\right]
        \log_2 2F(\bar{\mu}).
    \end{split}
\end{equation}
\end{widetext}

\subsection{Dimensionless parametrization}\label{sec:dimless_param}
\Cref{eqn:simple_fidelity,eqn:simple_ebr} can be further simplified through a dimensionless parametrization. We define the dimensionless flux $x$ and noise parameter $y_n$ as
\begin{equation}\label{eqn:dimless_params}
x\coloneqq\tau\bar{\mu} \;\;\;\;\;\; ; \;\;\;\;\;\;
y_n\coloneqq\frac{\tau d_n}{\eta_n}.
\end{equation}
The physical interpretation of $x$ is the mean number of biphotons produced in a coincidence window $\tau$, equivalent to a pair production probability per pulse---where the ``pulse'' is defined in an effective sense for this CW-pumped case. The noise parameter $y_n$ can be interpreted as a ratio of probabilities. The numerator is the probability of observing a noise event within a coincidence window $\tau$; the denominator is the probability of detecting a desired photon, given that it was produced within that same window.

After direct substitution of $x$ and $y$, \cref{eqn:simple_fidelity} becomes
\begin{equation}\label{eqn:dimless_fidelity}
    \begin{split}
        \mathcal{F}(x)&\coloneqq F\left(\frac{x}{\tau}\right)\\
        &=\frac{1}{4}\left[1+\frac{3x}{x^2+(2y_1+2y_2+1)x+4y_1y_2}\right]. 
    \end{split}
\end{equation}
 We then define the dimensionless EBR $\mathcal{R}$ as
\begin{equation}\label{eqn:dimless_ebr}
    \begin{split}
    & \mathcal{R}(x)\coloneqq\frac{\tau}{\eta_1\eta_2}R\left(\frac{x}{\tau}\right)\\
    & =\left[x^2+(2y_1+2y_2+1)x+4y_1y_2\right]\times\log_2 2\mathcal{F}(x).
    \end{split}
\end{equation}
Significantly, this parametrization reveals that the dark count rate and efficiency impact the functional form of fidelity and EBR only via their ratio, so that each user can be characterized by a single number $y_n$ that quantifies the quality of their transmission path and receiver. Moreover, although $\mathcal{F}$ and $\mathcal{R}$ are related to each other, 
they depend on flux $x$ in markedly different ways. Indeed, the interplay between fidelity and EBR, noted in previous quantum networking demonstrations~\cite{Alshowkan2021}, proved one of the key motivations for the current investigation and presents the quantum network engineer with non-trivial tradeoffs in evaluating competing allocations. 

\subsection{Model maxima}\label{sec:model_maxima}
\begin{figure*}[t!]
    \centering
    \includegraphics[width=\linewidth]{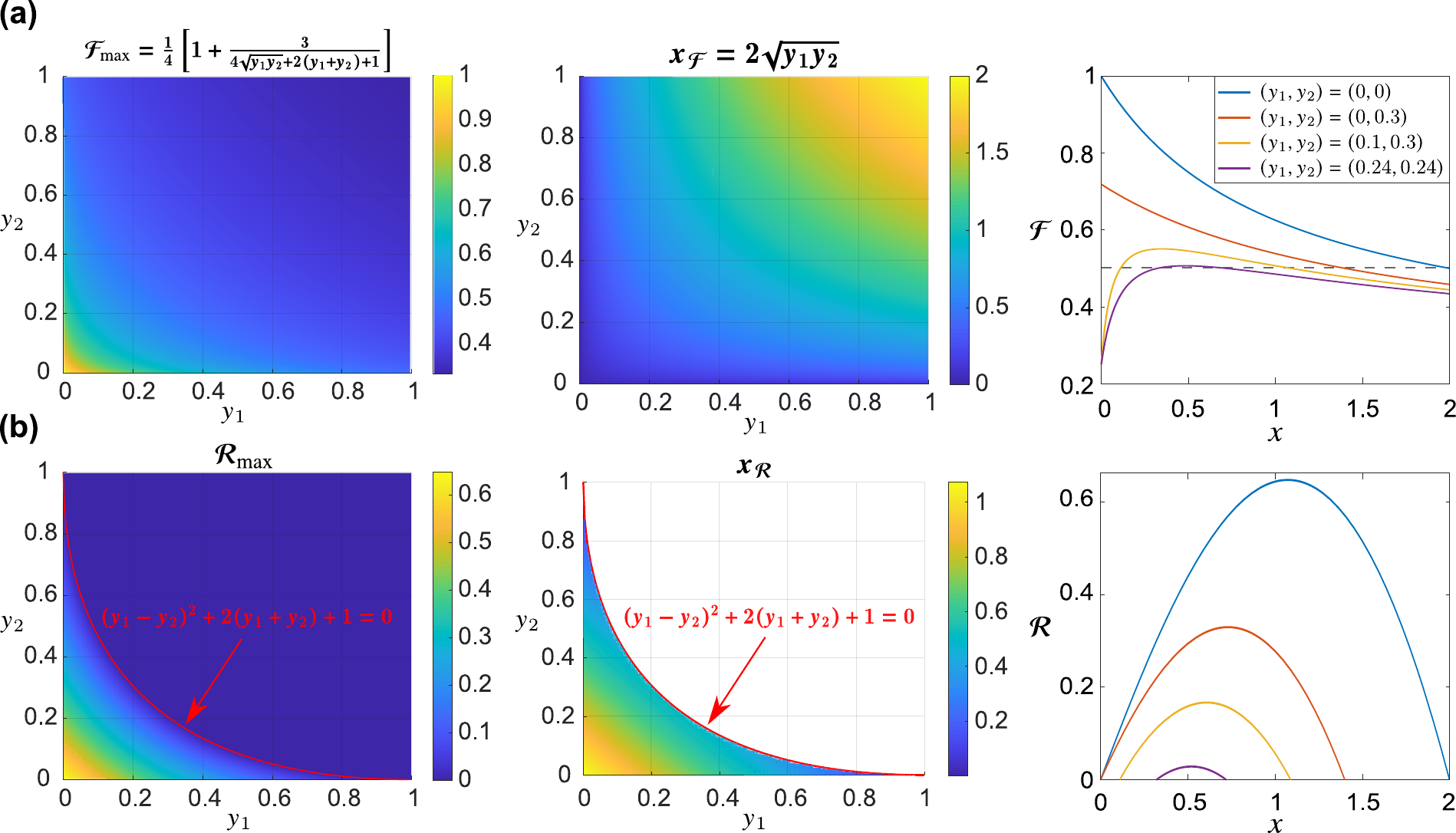}
    \phantomsubfloat{\label{fig:fid_optimal_surfs}}
    \phantomsubfloat{\label{fig:ebr_optimal_surfs}}
    \vspace{-2\baselineskip}
    \caption{(a) Surface plots in noise parameter space of maximum fidelity $\mathcal{F}_{\max}$ (left) and corresponding flux $x_\mathcal{F}$ (middle), as well as 
    full plots of $\mathcal{F}(x)$ for select noise parameter pairings (right). The dashed gray line marks $\mathcal{F}=0.5$, below which entanglement is lost for the Werner state in \cref{eqn:simple_model}. (b) Surface plots in noise parameter space of  maximum dimensionless EBR $\mathcal{R}_{\max}$ (left) and corresponding flux $x_\mathcal{R}$ (middle), along with $\mathcal{R}(x)$ for particular noise parameters (right). The boundary curve [\cref{eqn:boundary}] is drawn in red.} 
    \label{fig:analysis_curves_surfs}
\end{figure*}

Fidelity in \cref{eqn:dimless_fidelity} is an algebraic function of $x$ so that 
closed-form expressions for the location and value of the optimum 
can be found. Setting $\partial\mathcal{F}/\partial x =0$, the maximal fidelity is found to occur at 
%
$x_{\mathcal{F}}=2\sqrt{y_1y_2}$
%
with a corresponding maximum of
%
\begin{equation}\label{eqn:max_fidelity}
\begin{split}
    \mathcal{F}_{\max}(y_1,y_2)&=\mathcal{F}(x_\mathcal{F};y_1,y_2)\\&=\frac{1}{4}\left[1+\frac{3}{4\sqrt{y_1y_2}+2(y_1+y_2)+1}\right].
\end{split}
\end{equation}
\Cref{fig:fid_optimal_surfs} gives the surface plots of $x_\mathcal{F}$ and \cref{eqn:max_fidelity} as well as curves of \cref{eqn:dimless_fidelity} for select values of $y_1$ and $y_2$. The surface plots show that as $y_1$ or $y_2$ increases $\mathcal{F}_{\max}$ decreases and the location of $\mathcal{F}_{\max}$ shifts towards higher $x$. Eventually, $\mathcal{F}_{\max}$ approaches 0.5, below which the log-negativity vanishes and the state becomes separable~\cite{Horodecki1996}.


Unlike fidelity, the expression for EBR in \cref{eqn:dimless_ebr} and its first derivative with respect to $x$ are both transcendental. Thus, it is impossible to derive closed-form expressions for the location $x_\mathcal{R}$ and value $\mathcal{R}_{\max}$ of the EBR maximum. Nonetheless, boundaries can be derived within which $\mathcal{R}_{\max}>0$. From graphical experiments, it is known that $\partial^2\mathcal{R}/\partial x^2<0$ where $x\geq0$. Therefore, $\mathcal{R}_{\max}>0$ if and only if $\mathcal{R}$ has two nonnegative real roots, which arise from the logarithmic term and occur when the state is no longer entangled ($\mathcal{F}=1/2$). For the roots $z_{\mathcal{R}}=1-y_1-y_2\pm\sqrt{(y_1-y_2)^2-2(y_1+y_2)+1}$ to exist and be unique, we must have 
%
\begin{equation}\label{eqn:boundary}
    (y_1-y_2)^2-2(y_1+y_2)+1>0.
\end{equation}
%
Thus, users with noisy detectors may be compensated by users with low-noise detectors. For example, if $y_1=0.8$, then entanglement is still possible so long as $y_2<0.011$, albeit with a relatively low maximum EBR ($\mathcal{R}_{\max}=0.0255$).
Surfaces showing $\mathcal{R}_{\max}$ and $x_\mathcal{R}$ were generated using \verb+fminbnd()+ in MATLAB~R2021a and are plotted in \cref{fig:ebr_optimal_surfs}. Like fidelity, $\mathcal{R}_{\max}$ decreases with increasing $y_1$ or $y_2$. However, $x_\mathcal{R}$ decreases as either $y_1$ or $y_2$ increases. Once the condition in \cref{eqn:boundary} is no longer satisfied, $\mathcal{R}=0~\forall~x$, and the location of a maximum can no longer be defined. Also note that the largest dimensionless EBR possible ($\mathcal{R}_{\max}$ when $y_1=y_2=0$) is 0.6475, a number which follows from the definitions and flux dependencies in \cref{eqn:dimless_ebr} but does not seem to possess any intuitive significance.

The analysis above establishes fundamental limits for the entangled quantum state shared by two users: given channel losses and dark count rates, reflected in parameters $y_1$ and $y_2$, $\mathcal{F}_{\max}$ and $\mathcal{R}_{\max}$ follow immediately, providing upper bounds that no provisioning scenario can exceed. The extent to which these maxima can be approached for a given link thus depends on the available resources (flux and wavelength channels) and the particular fitness function selected for optimization. Intuitively, the optimal flux for fidelity $x_\mathcal{F}$ balances the effects of noise from dark counts (which increases with smaller $x$) and multi-pair emission (which grows with increasing $x$). On the other hand, the larger EBR optimum ($x_\mathcal{R}>x_\mathcal{F}$ whenever $\mathcal{R}_{\max}>0$)  results from balancing the tradeoff between state quality (maximized at $x_\mathcal{F}$) and detection rate (maximized as $x\rightarrow\infty$); this feature of EBR lends credence to our arguments for it as an application-agnostic metric for quantum networking~\cite{Alshowkan2021, Alshowkan2021b}. Nevertheless, as we discuss later, fidelity is likely to remain significant in its own right in near-term quantum networks.

Finally, as an aside, the conclusions found in this section follow logically from previous models such as~\cite{Takesue2010}, yet to our knowledge no explicit derivation has been provided in the literature. Accordingly, these findings offer useful design considerations  more generally for any SPDC entanglement source, regardless of application. When all experimentally controllable nonidealities are removed (mode mismatch, synchronization, etc.), these limits remain due to channel and detector properties and the fundamental physics of probabilistic pair production.

\section{Network optimization}\label{sec:network_opt}
\subsection{Problem formulation \& complexity}\label{sec:prob_form}
Using the model given in \cref{sec:model}, we now introduce the entangled flux allocation (EFA) problem, which can be formulated as a nonlinear integer optimization problem. Taking as given $K$ energy-correlated pairs of frequency bands,
the objective of the EFA problem is to find an allocation of the $K$ channels that optimizes a desired entanglement network performance metric. We can represent an allocation with the vector $\bm{\alpha}\in\{0,1,2,\dots,L\}^{K}$ where $\alpha_k=l$ if the $k$th channel pair is assigned to link $l$. If $\alpha_k=0$, then channel $k$ is allocated to the ``reserve link,'' a virtual link that holds the flux of any unallocated channels---i.e., channels which are sent to no network users. Assuming each of the $K$ channels is unique, there are $(L+1)^K$ unique channel allocations.

The complexity of EFA can be determined through comparison with the multiple subset sum problem (MSSP). In the formal definition of MSSP, the objective is to partition a set of $N$ weighted items into $M$ bins each with positive integer capacity $c$, ``filling'' the bins as much as possible without exceeding capacity $c$~\cite{Caprara2000}. If the integer constraints on the weights and capacity are lifted, the $N$ items, $N$ weights, and $M$ bins of capacity $c$ of MSSP are respectively analogous to the $K$ frequency channels, $K$ channel fluxes, and $L$ entanglement links with target flux $\phi_l$ of our simplified EFA model. Whereas all $M$ bins in MSSP have the same capacity $c$, each link may have a distinct $\phi_l$, the exact value of which may be calculated from \cref{eqn:dimless_fidelity} or \cref{eqn:dimless_ebr}. Furthermore, whereas MSSP forbids bin loads from exceeding capacity $c$, link fluxes are allowed to exceed $\phi_l$ in EFA. Because EFA is obtained by relaxing some constraints of the MSSP, it is a superproblem of MSSP. Further, the formal MSSP is known to be strongly NP-hard~\cite{Caprara2000}. Hence, the complexity of the simplified EFA is at least strongly NP-hard.

As another connection to previous work, we note here that the EFA problem bears some resemblance to the routing and spectrum allocation (RSA) problem in classical flex-grid networks. The objective of RSA is to efficiently utilize a spectrum of frequencies to satisfy traffic demands between source and destination nodes of a physical network. 
Low-bandwidth demands can be assigned an appropriately smaller portion of the spectrum, effectively increasing network capacity. The RSA problem is generally NP-hard \cite{Klinkowski2011,Busing2017} and is typically solved via some form of integer programming \cite{Christodoulopoulos2011,Klinkowski2011,Ruiz2013} or meta-heuristic algorithms \cite{Wang2013,Klinkowski2013,Lezama2016,Markovic2017}. 

In RSA, a route through the physical network must be found in coordination with the assignment of  frequency channels. Applications often enforce the continuity and contiguity constraints \cite{Velasco2017}. The first ensures that frequency channels assigned at the source are not used by nodes en route to the destination. The second ensures that multiple spectral slices allocated to a demand are adjacent in the frequency domain. In contrast, such constraints can be relaxed in the EFA problem considered here. The assumption of a star physical topology whereby each node possesses a direct fiber connection to the source (cf. \cref{fig:qlan_layers}) ensures that any allocation automatically satisfies the continuity requirement.
The reliance on frequency correlations resulting from broadband SPDC means that distributing discontinuous spectra to each user poses no fundamental problem in EFA and, in fact, has been considered experimentally~\cite{Lingaraju2021}. 

However, in more complex entanglement distribution networks (e.g., with multiple SPDC sources, nested WSSs, or multi-hop links), the explicit consideration of a continuity constraint will be required in EFA as well. Moreover, spectral contiguity---even if not intrinsically necessary for EFA---will likely prove practically important in order to minimize the effects of wavelength-dependent fiber birefringence; the greater the separation between spectral slices $k$, the less likely their polarization rotations $U_k$ can be compensated in tandem by a single operation.

Finally, the fitness function used in RSA is distinct from that used here. In RSA, the objective is to minimize spectrum utilization while still satisfying requests. The corresponding fitness function is often formulated as a single integer denoting the highest index of spectral slices used. Compare this with the fitness function in this paper, which makes use of the nonlinear functions \cref{eqn:dimless_fidelity,eqn:dimless_ebr} to determine entanglement quality.

\subsection{Genetic algorithm}\label{sec:ga}
To solve the NP-hard EFA problem, we turn to evolutionary algorithms (EA), which use metaheuristic search strategies modeled after biological processes to find optimal solutions to difficult problems. Since these strategies are independent of the problem itself, they can be applied regardless of problem size, linearity, or availability of gradients. Well-known examples of EAs include ant colony optimization (ACO)~\cite{Dorigo1996,Dorigo1992}, particle swarm optimization (PSO)~\cite{Eberhart1995}, and GAs~\cite{Holland1992}. EAs have already found applications in tackling problems in quantum communications~\cite{Gyongyosi2019,Krastanov2019,FerreiradaSilva2021}.

\begin{figure}[tb]
    \centering
    \includegraphics[width=1\linewidth]{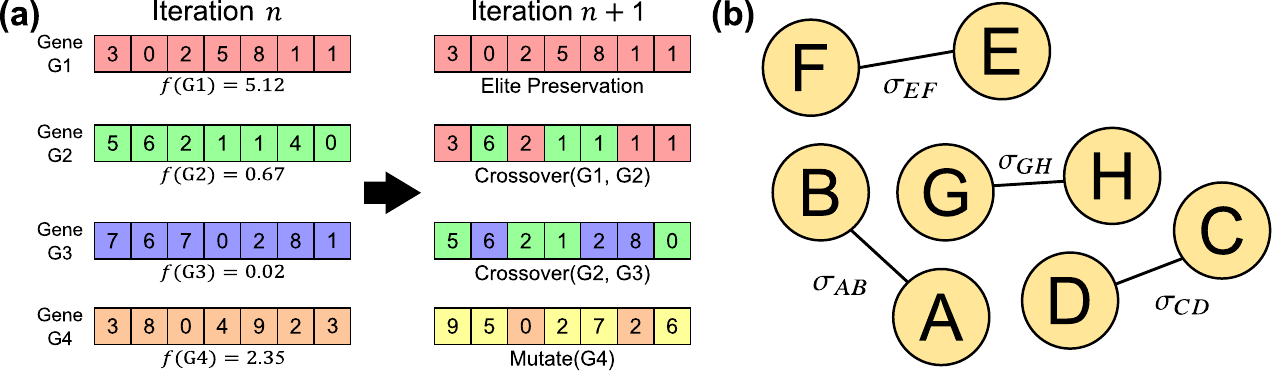}
    \phantomsubfloat{\label{fig:ga_example}}
    \phantomsubfloat{\label{fig:opt_net}}
    \vspace{-2\baselineskip}
    \caption{(a) GA example. Each gene of the population represents a possible solution and is evaluated using a fitness function $f$. The genes of the next iteration are generated by three methods: (1) preservation, (2) crossover, and (3) mutation. Elite genes (i.e., genes with highest $f$) are preserved and survive to the next generation. In crossover, children are constructed by selecting two parents and combining the genes in some fashion, usually by randomly selecting elements. In mutation, elements of a gene are randomly changed to new values. (b) Network topology for optimization Scenarios 1--3. In Scenario 4, seven additional one-to-one links are included.}
    \label{fig:ga_example_network}
\end{figure}

We apply the GA as implemented in MATLAB R2021a~\cite{Mathworks} to tackle the EFA problem. A GA is a stochastic optimization algorithm that maximizes (or minimizes) a so-called fitness (or objective) function by mimicking biological evolution. It begins with a randomly generated population of genes, all of which represent inputs to a user-defined fitness function. The genes that maximize the fitness survive to the next iteration unaltered. A new population is created by randomly crossing over (i.e., combining) or mutating existing genes. These processes are illustrated in \cref{fig:ga_example}. A GA is desirable because of its flexibility and adaptability to continuous, integer, or mixed-integer problems of any size. Although discrete versions of ACO \cite{Dorigo1999,Schluter2009} and PSO \cite{Kennedy1997,Afshinmanesh2008,Alnas2021} exist, they are more amenable to binary problems as opposed to integer problems such as EFA. In our optimizations, we use a population of 200 genes and a crossover fraction of 0.8. The optimization terminates when the maximum fitness achieved does not change for 100 consecutive iterations.
%

We apply this GA in four different optimization scenarios motivated by realistic network conditions to test its tractability for the EFA problem. Each scenario uses the following fitness function:
%
\begin{subequations}\label{eqn:fitness_func}
    \begin{equation}
        f=\sum_{l=1}^L\beta_l(\mathcal{R}_l;\mathcal{F}_l,\mathcal{F}_{\min})
    \end{equation}
where 
%
    \begin{equation}\label{eqn:fitness_sub}
        \beta_l(\mathcal{R}_l;\mathcal{F}_l,\mathcal{F}_{\min})=
        \begin{cases}
            \frac{\mathcal{R}_l}{\mathcal{R}_{l,\max}}, & \mathcal{F}_l\geq\mathcal{F}_{\min}\\
            -1, & \text{otherwise}
        \end{cases}
    \end{equation}
\end{subequations}%
and $\mathcal{R}_l/\mathcal{R}_{l,\max}$ is the link $l$ EBR normalized to the link's theoretical maximum EBR. \Cref{eqn:fitness_func} is designed to maximize each link's EBR while maintaining a fidelity of at least $\mathcal{F}_{\min}$ without enforcing an explicit constraint that would add significant computational complications. For benchmarking purposes, we 
define $f_\infty$ as the best possible fitness given constraints and access to infinite resources (e.g., unlimited total flux and channels $K$)

\Cref{tab:scenarios} provides the values $L$, $K$, $\mathcal{F}_{\min}$, and $f_\infty$  for each of the four scenarios examined. \Cref{fig:opt_net} shows the five-link, one-to-one entanglement network used in Scenarios 1--3. Scenario 4 uses a larger, 12-link one-to-one network. For each scenario, optimizations are run with an increasing number of channels $K$. As GAs are stochastic, five independent optimizations are performed for each $K$, and the result with the largest fitness is kept.
In addition to flux allocation, we include the total biphoton flux $\mu_\mathrm{tot}$ as an optimization variable. Changing this is analogous to changing the pump power of the central provider's entangled light source. Although our algorithm is designed to handle variable channel fluxes, in the scenarios here we make the simplification that the total flux is uniformly distributed across the $K$ available channels: $\mu_k=\mu_\mathrm{tot}/K$. Note that this uniform distribution results in redundant allocations. Under such conditions, it does not matter \textit{which} channels are allocated to a link, only \textit{how many} channels are allocated. Thus the number of unique allocations reduces from $(L+1)^K$ to $\binom{K+L}{L}$.
%
%
\begin{table}[b!]
\centering
\caption{Parameters for the four optimization scenarios. Given the minimum fidelity threshold $\mathcal{F}_{\min}$, the best achievable fitness is calculated and reported as $f_\infty$.}
\label{tab:scenarios}
\begin{ruledtabular}
\begin{tabular}{ccccccc}
Scenario & $L$ & $K$ & $\mathcal{F}_{\min}$ & \multicolumn{1}{l}{$f_\infty$} \\ 
\hline
1 & 5  & \{5,~10,~20,~40\}  & 0   & 5    \\ 
2 & 5  & \{5,~10,~20,~40\}  & 0.7 & 3.39 \\ 
3 & 5  & \{5,~10,~20,~40\}  & 0.9 & 0.91 \\ 
4 & 12 & \{12,~24,~48,~96\} & 0.7 & 7.9  \\  
\end{tabular}
\end{ruledtabular}
\end{table}
%
\begin{table*}[t]
\centering
\caption{Noise parameter pairings and maximum fidelity $\mathcal{F}_{\max}$ for each link in each scenario. Raised dots are shown for links excluded from a scenario.}
\label{tab:noise_params}
\begin{ruledtabular}
\begin{tabular}{cllllllllllll}
\multirow{2}{*}{Link} & \multicolumn{3}{c}{Scenario 1} & \multicolumn{3}{c}{Scenario 2} & \multicolumn{3}{c}{Scenario 3} & \multicolumn{3}{c}{Scenario 4} \\ \cline{2-13} 
 & \multicolumn{1}{c}{$y_1$} & \multicolumn{1}{c}{$y_2$} & \multicolumn{1}{c}{$\mathcal{F}_{\max}$} & \multicolumn{1}{c}{$y_1$} & \multicolumn{1}{c}{$y_2$} & \multicolumn{1}{c}{$\mathcal{F}_{\max}$} & \multicolumn{1}{c}{$y_1$} & \multicolumn{1}{c}{$y_2$} & \multicolumn{1}{c}{$\mathcal{F}_{\max}$} & \multicolumn{1}{c}{$y_1$} & \multicolumn{1}{c}{$y_2$} & \multicolumn{1}{c}{$\mathcal{F}_{\max}$} \\ \hline
AB & 0 & 0 & 1 & 0 & 0 & 1 & 0 & 0 & 1 & 0 & 0 & 1 \\
CD & 0.04 & 0.007 & 0.90 & 0.04 & 0.007 & 0.90 & 0.0034 & 0.006 & 0.99 & 0.0034 & 0.0006 & 0.99 \\
EF & 0 & 0.125 & 0.85 & 0 & 0.125 & 0.85 & 0.0104 & 0.0018 & 0.97 & 0 & 0.0357 & 0.95 \\
GH & 0.11 & 0.019 & 0.77 & 0.11 & 0.019 & 0.77 & 0.0179 & 0.0031 & 0.95 & 0.0299 & 0.0051 & 0.92 \\
IJ & 0.15 & 0.025 & 0.72 & 0.15 & 0.025 & 0.72 & 0 & 0.0515 & 0.93 & 0.0385 & 0.0066 & 0.90 \\
KL & $\cdots$ & $\cdots$ & $\cdots$ & $\cdots$ & $\cdots$ & $\cdots$ & $\cdots$ & $\cdots$ & $\cdots$ & 0.0625 & 0.0107 & 0.85 \\
MN & $\cdots$ & $\cdots$ & $\cdots$ & $\cdots$ & $\cdots$ & $\cdots$ & $\cdots$ & $\cdots$ & $\cdots$ & 0.0733 & 0.0126 & 0.83 \\
OP & $\cdots$ & $\cdots$ & $\cdots$ & $\cdots$ & $\cdots$ & $\cdots$ & $\cdots$ & $\cdots$ & $\cdots$ & 0 & 0.1818 & 0.80 \\
QR & $\cdots$ & $\cdots$ & $\cdots$ & $\cdots$ & $\cdots$ & $\cdots$ & $\cdots$ & $\cdots$ & $\cdots$ & 0.1106 & 0.019 & 0.77 \\
ST & $\cdots$ & $\cdots$ & $\cdots$ & $\cdots$ & $\cdots$ & $\cdots$ & $\cdots$ & $\cdots$ & $\cdots$ & 0.125 & 0.0214 & 0.75 \\
UV & $\cdots$ & $\cdots$ & $\cdots$ & $\cdots$ & $\cdots$ & $\cdots$ & $\cdots$ & $\cdots$ & $\cdots$ & 0.1489 & 0.0256 & 0.72 \\
WX & $\cdots$ & $\cdots$ & $\cdots$ & $\cdots$ & $\cdots$ & $\cdots$ & $\cdots$ & $\cdots$ & $\cdots$ & 0 & 0.2979 & 0.72 \\
\end{tabular}%
\end{ruledtabular}
\end{table*}

\subsection{Optimization results}\label{sec:opto_results}
\begin{figure}[b!]
    \centering
    \includegraphics[width=\linewidth]{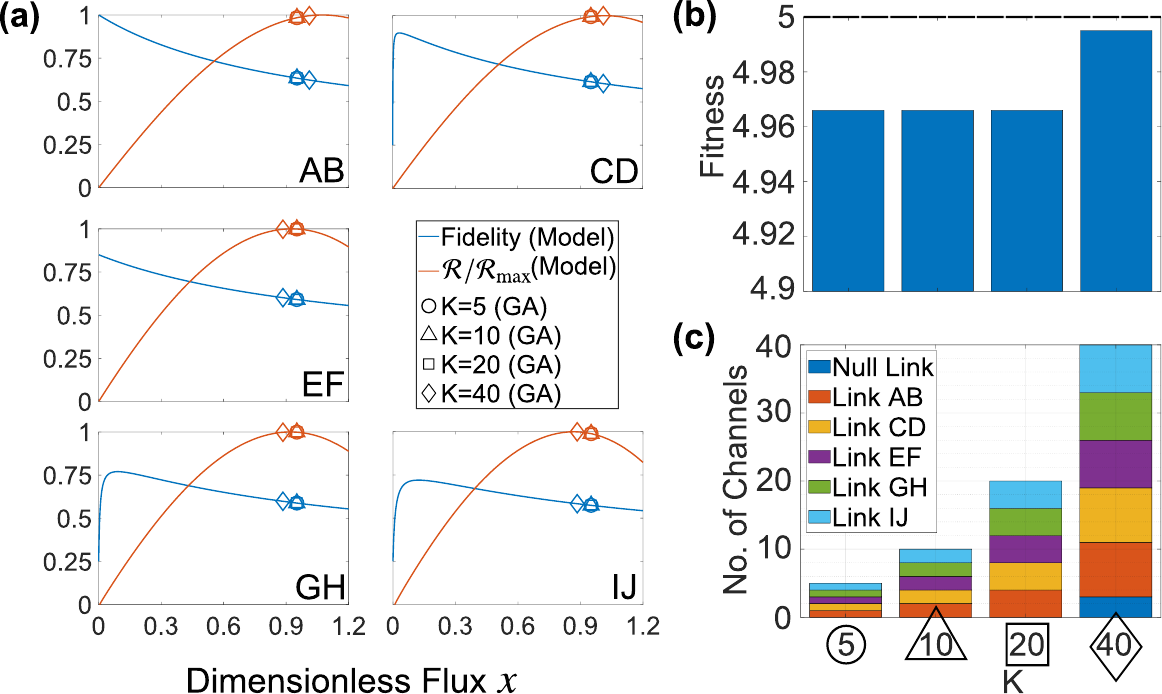}
    \phantomsubfloat{\label{fig:opt1_curves}}
    \phantomsubfloat{\label{fig:opt1_fitness}}
    \phantomsubfloat{\label{fig:opt1_alloc}}
    \vspace{-2\baselineskip}
    \caption{Scenario 1 optimization results. A five-link entanglement network is optimized to maximize the EBR of all five links without regard to link fidelity. (a)~GA-achieved fidelity and EBR along model curves [Eqs.~(\ref{eqn:dimless_fidelity},\ref{eqn:dimless_ebr})]. (b) Best fitness achieved for each $K$, compared to the best possible given fidelity constraints and sufficient resources of $f_\infty=5$. (c) Distribution of frequency channels to links for each $K$.}
    \label{fig:opt1_results}
\end{figure}

\begin{figure}[b!]
    \centering
    \includegraphics[width=\linewidth]{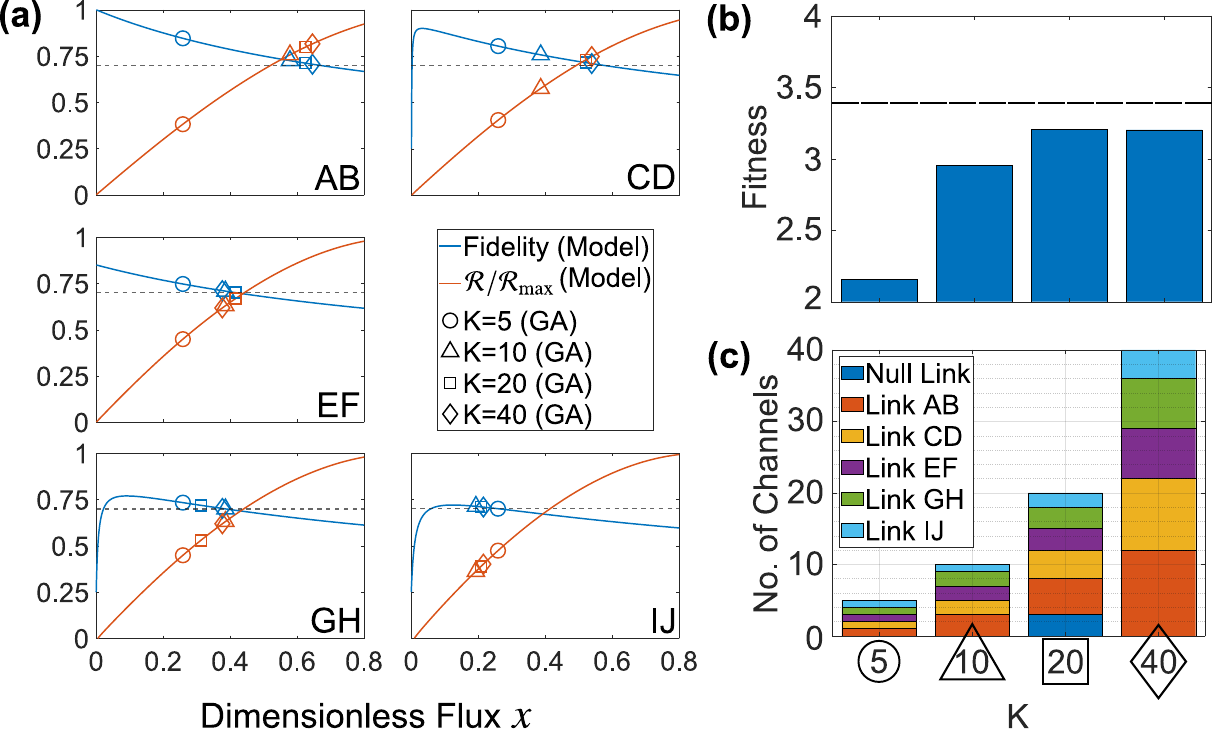}
    \phantomsubfloat{\label{fig:opt2_curves}}
    \phantomsubfloat{\label{fig:opt2_fitness}}
    \phantomsubfloat{\label{fig:opt2_alloc}}
    \vspace{-2\baselineskip}
    \caption{Scenario 2 optimization results. A five-link entanglement network is optimized to maximize the EBR of all five links while maintaining $\mathcal{F}_l\geq0.7$. (a)~GA-achieved fidelity and EBR along model curves [Eqs.~(\ref{eqn:dimless_fidelity},\ref{eqn:dimless_ebr})]. (b) Best fitness achieved for each $K$, compared to the best possible given fidelity constraints and sufficient resources of $f_\infty=3.39$. (c) Distribution of frequency channels to links for each $K$.}
    \label{fig:opt2_results}
\end{figure}

\begin{figure}[tb!]
    \centering
    \includegraphics[width=\linewidth]{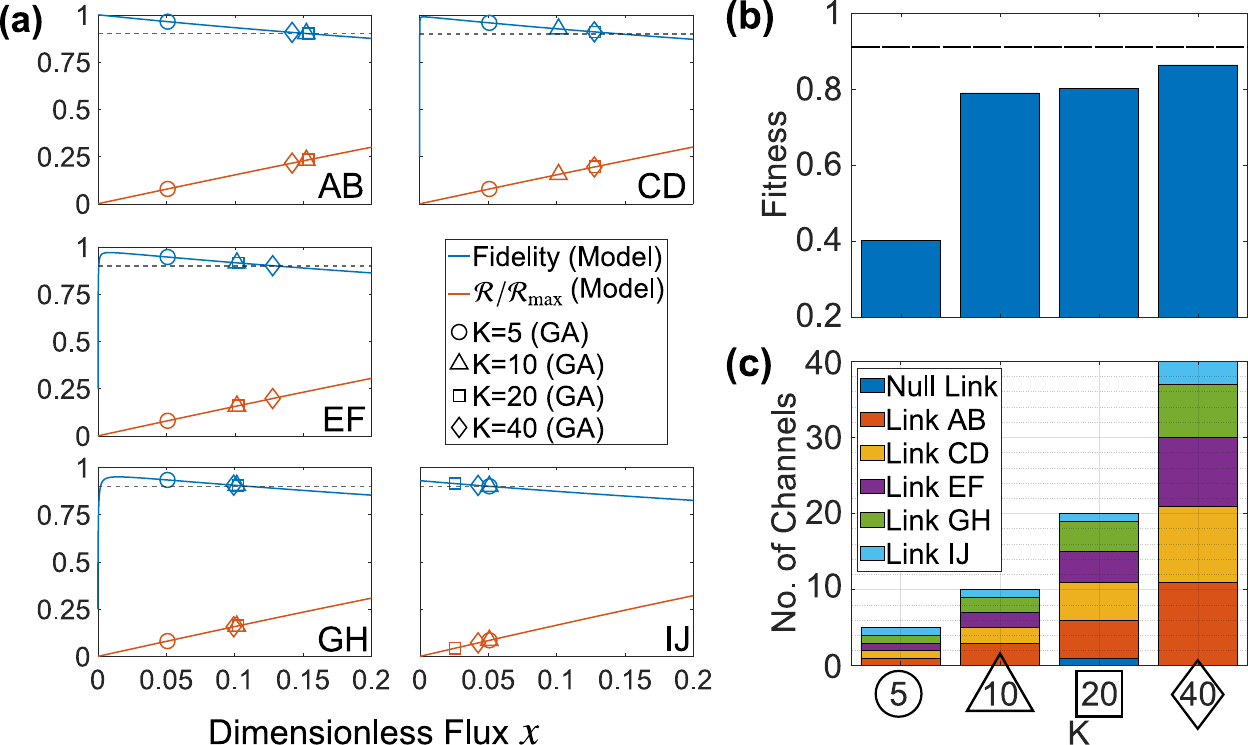}
    \phantomsubfloat{\label{fig:opt3_curves}}
    \phantomsubfloat{\label{fig:opt3_fitness}}
    \phantomsubfloat{\label{fig:opt3_alloc}}
    \vspace{-2\baselineskip}
    \caption{Scenario 3 optimization results. A five-link entanglement network is optimized to maximize the EBR of all five links while maintaining $\mathcal{F}_l\geq0.9$. (a)~GA-achieved fidelity and EBR along model curves [Eqs.~(\ref{eqn:dimless_fidelity},\ref{eqn:dimless_ebr})]. (b) Best fitness achieved for each $K$, compared to the best possible given fidelity constraints and sufficient resources of $f_\infty=0.91$. (c) Distribution of frequency channels to links for each $K$.}
    \label{fig:opt3_results}
\end{figure}

\begin{figure}[bt!]
    \centering
    \includegraphics[width=\linewidth]{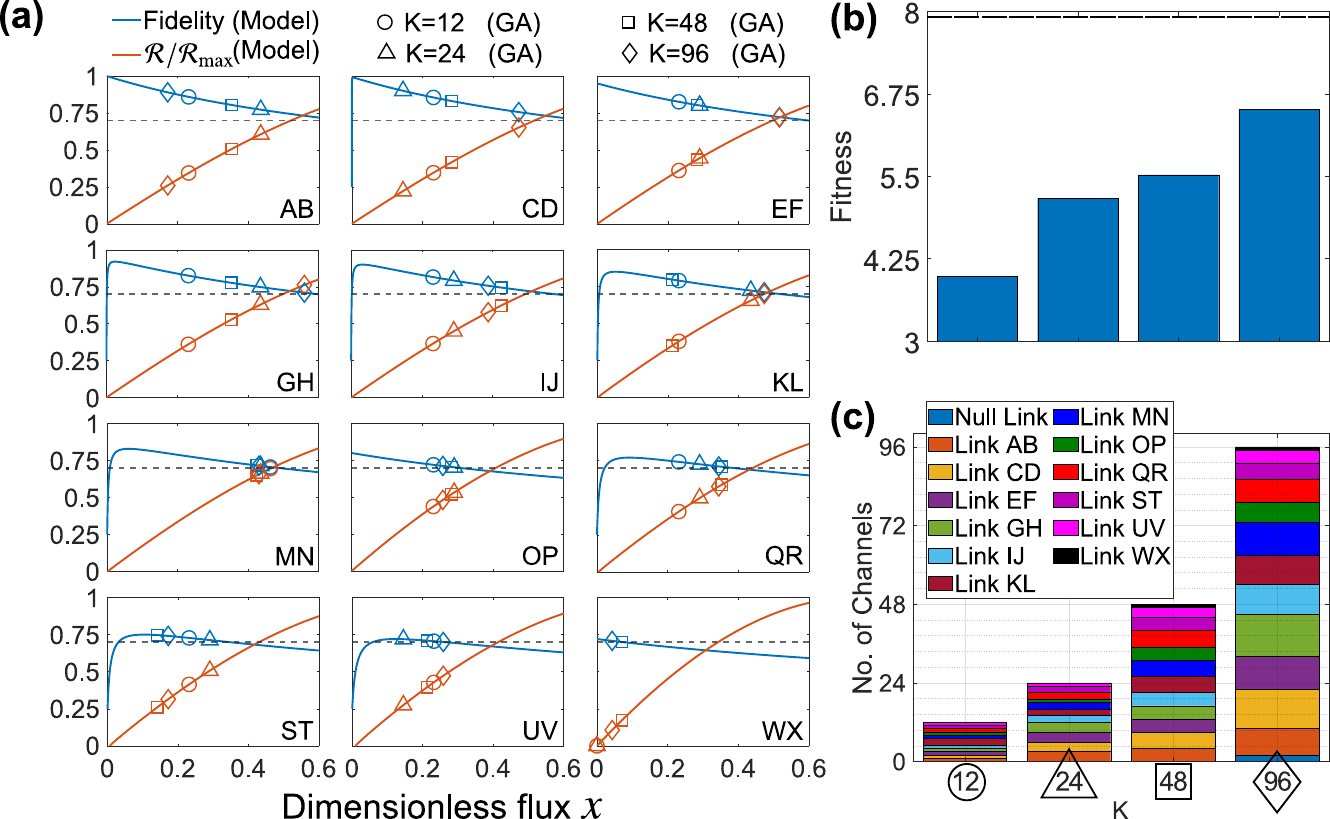}
    \phantomsubfloat{\label{fig:opt4_curves}}
    \phantomsubfloat{\label{fig:opt4_fitness}}
    \phantomsubfloat{\label{fig:opt4_alloc}}
    \vspace{-2\baselineskip}
    \caption{Scenario 4 optimization results. A 12-link entanglement network is optimized to maximize the EBR of all 12 links while maintaining $\mathcal{F}_l\geq0.7$ for all 12 links. (a)~GA-achieved fidelity and EBR along model curves [Eqs.~(\ref{eqn:dimless_fidelity},\ref{eqn:dimless_ebr})]. (b) Best fitness achieved for each $K$, compared to the best possible given fidelity constraints and sufficient resources of $f_\infty=7.9$. (c) Distribution of frequency channels to links for each $K$.}
    \label{fig:opt4_results}
\end{figure}

Using the model in \cref{sec:model}, three networks were designed for use in four optimization scenarios. 
\Cref{tab:noise_params} gives the noise parameters and the maximum achievable fidelities for each link for each scenario. As context for these values, for $\tau=1$~ns we estimate that the superconducting nanowire detector (SNSPD) used by Alice in the experiment of \cref{sec:exp} below corresponds to $y_A\approx1.3\times 10^{-7}$ ($y_A\approx8.3\times10^{-6}$) with extra channel losses neglected (included); for Bob, who utilizes an InGaAs avalanche photodiode (APD), $y_B\approx3.5\times 10^{-5}$ ($y_B\approx1.7\times10^{-2}$) without (with) additional channel losses. In order to explore the impact of heterogeneous nodes on the EFA problem, in \Cref{tab:noise_params} we have intentionally selected a much broader range of noise parameters than these experimental examples, leading to links with a wide spread in maximum fidelity $\mathcal{F}_{\max}$. For Scenarios 1, 2, and 4, all links lie in the interval $\mathcal{F}_{\max}\in[0.72, 1]$; for Scenario 3, $\mathcal{F}_{\max}\in[0.93, 1]$. In every case, $\mathcal{F}_{l,\max}>0.5$ so that $\mathcal{R}>0$ is possible for each pair of examined nodes. Final optimization results are given in Figs.~\ref{fig:opt1_results}--\ref{fig:opt4_results}. For each scenario, three types of plots are provided: (a) fidelity and normalized EBR achieved via GA optimization, plotted as points on the curves of $\mathcal{F}_l(x_l)$ and $\mathcal{R}_l(x_l)/\mathcal{R}_{l,\max}$; (b) fitness $f$ for each $K$ as a bar graph with the maximum achievable fitness $f_\infty$ marked as a dashed line; and (c) number of channels allocated to each link for each $K$ plotted as a stacked bar graph. 

The optimization goal of Scenario 1 is to maximize the EBR of all five network links without any fidelity constraints [$\mathcal{F}_{\min}=0$ in \cref{eqn:fitness_func}]. 
The best achievable solution has fitness $f_\infty=5$, which corresponds to every link achieving maximum normalized EBR. \Cref{fig:opt1_results} shows that this goal is closely met for any $K$. For $K<40$, GA finds allocations with $f=4.97$, a 0.68\% deviation from $f_\infty=5$. This deviation decreases to 0.1\% with $K=40$ where an allocation with $f=4.995$ is found. The fact that the EBRs for all links are maximized by similar flux values facilitates this performance. Indeed, for $K\in\{5,10,20\}$, the optimal allocations divide the channels evenly between all links, and so the fitness obtained in all cases remains approximately the same. It is only at $K=40$ where the increased spectral granularity leads to a nonuniform provisioning as optimal, with AB and CD receiving 8 channels each and EF, GH, and IJ receiving 7. 
%

In Scenario 1, we concentrated on optimizing EBR only, a promising candidate in our view for a universal metric of an entangled link's quality. Nonetheless, utilizing this entanglement in practice would demand implementation of entanglement distillation~\cite{Bennett1996,Asano2015,Kalb2017,Chen2020}, an extremely demanding protocol that---although critical for quantum networking in the long-term---is beyond the capabilities of many existing quantum network testbeds. Consequently, users on near-term quantum networks will likely be expected to request a threshold state quality defined by the particular application, so that optimization with $\mathcal{F}_{\min}$ specified should be used to determine a bandwidth allocation.
Thus, in Scenario 2, the same network as in Scenario 1 was optimized, but now with a fidelity threshold. In order to ensure that all links have the potential to exceed the threshold, we must have $\mathcal{F}_{\min}\leq\min_{l}(\mathcal{F}_{l,\max})$. For Scenario 2, $\mathcal{F}_{\min}$ must therefore be less than 0.72, and so we select ${\mathcal{F}_{\min}=0.7}$ for numerical optimization. \Cref{fig:analysis_curves_surfs} shows that in the low-flux regime, EBR increases as fidelity decreases. Hence, the best achievable EBR given a fidelity threshold $\mathcal{F}_{\min}$ will have $\mathcal{F}_l=\mathcal{F}_{\min}$ for all links. For Scenario 2, this corresponds to an ideal fitness value of $f_\infty=3.39$. The introduction of a fidelity penalty  makes the problem more difficult to solve in comparison to Scenario 1. The best fitness achieved was $f=3.21$ (5.37\% deviation) at $K=20$. Interestingly, a slightly lower fitness of $f=3.20$ (5.56\% deviation) was achieved with $K=40$; since fitness should never decrease with $K$ (as allocations at smaller $K$ are a subset of those possible with larger $K$), this slight decrease is a computational artifact reflecting the growing numerical difficulties with constraints and increased resources.
%

More stringent fidelity constraints are enforced in Scenario 3. \Cref{eqn:fitness_func} is used with $\mathcal{F}_{\min}=0.9$, and the link noise parameters are chosen such that the maximum possible link fidelities are 1, 0.99, 0.97, 0.95, and 0.93 for AB, CD, EF, GH, and IJ, respectively. The best achievable solution has $\mathcal{F}_l=0.9$ for all links and has a corresponding fitness of $f_\infty=0.911$. \Cref{fig:opt3_results} shows that all links satisfy fidelity constraints for all $K$. For higher $K$, all links approach the ideal solution. The best fitness achieved is $f=0.864$ (5.14\% deviation).
%

Scenario 4 tests the capability of the GA when the network size is increased significantly. A 12-link network was designed such that $\mathcal{F}_{\max}\geq0.72$ for all links [according to the bound in \cref{eqn:max_fidelity}]. This network was optimized using $\mathcal{F}_{\min}=0.7$ in \cref{eqn:fitness_func}. Once again, the best achievable solution here has $\mathcal{F}_l=0.7$ for all links, corresponding to a maximum fitness of $f_\infty=7.9$. \Cref{fig:opt4_results} shows that all links satisfy the fidelity constraints for $K\geq48$. 
The best achieved fitness improved with available resources, ranging from $f=3.98$ (49.7\% deviation) for $K=12$ to $f=6.51$ (17.8\% deviation) for $K=96$.
%

Beyond limited computational resources, a source of difficulty in these scenarios arises whenever there exists a wide imbalance of noise parameters across the nodes, which challenges the source's ability to simultaneously satisfy the full range of optimal link demands at low channel numbers, i.e. $K\sim \mathcal{O}(L)$.
For example, for $K=5$ in Scenarios 1--3, all links were allocated one channel and thus received the same amount of flux. Ideally, to maximize the fitness function, $\mu_\mathrm{tot}$ (and thus $x_k=\tau\mu_k$) should be chosen such that the constraints for all links are simultaneously satisfied. Yet, for the $\mathcal{F}_{\max}$ settings of Scenarios 2 and 3, the low-noise links like AB require significantly more flux to reach their fidelity-constrained optimal EBR rates than noisy links such as link IJ. This noise imbalance means that $x_k$ cannot exceed the maximum flux that satisfies the noisiest link (or else violate the minimum fidelity condition). Mathematically speaking, we must have $x_k\leq\min_{l}(\phi_l)$ where $\phi_l$ is the link flux that optimizes the contribution of link $l$ to the fitness: $\phi_l=\mathrm{argmax}_{x\in\mathcal{S}}\,(\mathcal{R}_l(x))$, where $\mathcal{S}=\{x\in\mathbb{R}_{\geq 0}: \mathcal{F}_l(x)\geq\mathcal{F}_{\min}\}$.
Once $K$ increases, more channels can be allocated to links like AB and CD to compensate for the limited $x_k$. This is reflected in the (c) panels of Figs.~\ref{fig:opt2_results} and \ref{fig:opt3_results}, which show that the low-noise links receive proportionally more channels as $K$ increases.

These same considerations 
lead to interesting results in Scenario 4. \Cref{fig:opt4_results} shows that for $K\in\{12,24\}$, higher fitness is achieved by allocating \emph{no} channels to link WX, resulting in an undefined fidelity, $\mathcal{R}_{WX}=0$, and leaving more flux for lower-noise links. 
Because Scenario 4 has 12 links contributing to the fitness function as opposed to five in the previous three scenarios, the negative effect of limiting $x_k\leq\min_{l}(\phi_l)$ for 11 of the 12 links outweighs the benefit of satisfying the fidelity constraint for the noisiest link, WX. With $K\in\{48,96\}$, sufficient granularity in flux allocation is achieved, and it becomes possible to satisfy the fidelity threshold on all links simultaneously.

\section{Experimental test}\label{sec:exp}
The entanglement distribution model we have formulated here relies on the physical noise model for a single link as expressed by \cref{eqn:dimless_fidelity,eqn:dimless_ebr}. In order to explore the applicability of this model in a practical context, we perform experimental entanglement distribution tests in a deployed QLAN. Described in detail in~\cite{Alshowkan2021}, our QLAN consists of nodes in three separate buildings on the Oak Ridge National Laboratory campus and utilizes the flex-grid paradigm for distribution of polarization-entangled states. For the experiments here, we focus on the Alice--Bob (AB) link and perform quantum state tomography at a variety of pump laser powers in order to map the fidelity and EBR scaling behavior.

In order to minimize spectrally dependent birefringence effects, we utilize a single channel: Ch.~1 from \cite{Alshowkan2021}, which provides a 25~GHz-wide slice centered at 192.325~THz (192.300~THz) to Alice (Bob). At each pump laser power, we measure the received photons in the rectilinear ($H/V$) and diagonal ($D/A$) polarization bases, utilizing Bayesian inference to perform full tomography~\cite{Blume2010, Lukens2020b}. Refining the previous Bayesian procedure applied to the QLAN~\cite{Alshowkan2021,Alshowkan2021b}, we now take a Bures prior and assume a Poissonian likelihood, which better reflects the physical situation under test;
we point the reader to \cite{Lu2021} for details on this model for Bayesian inference.

\begin{figure*}[bt!]
\centering
\includegraphics[width=0.9\linewidth]{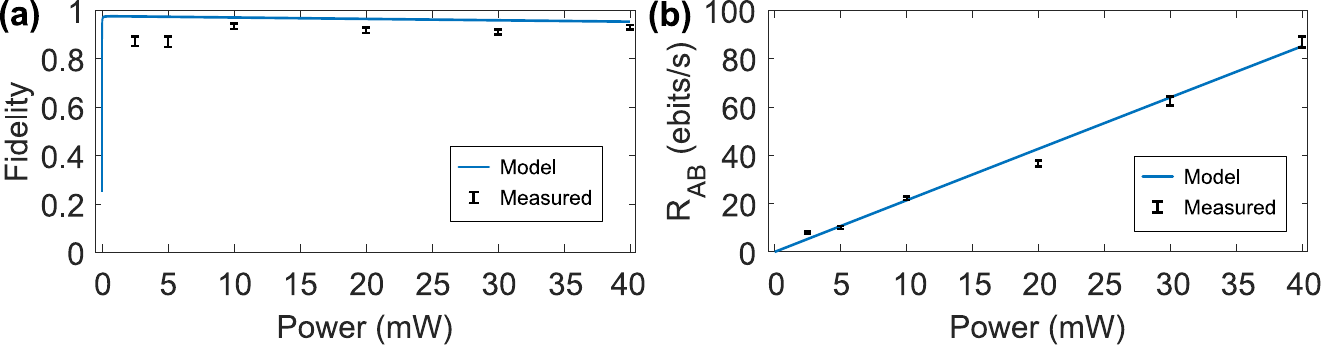}
\caption{\label{fig:exp} Experimental tests on a deployed QLAN. (a)~Fidelity and (b)~dimensioned EBR $R_\mathrm{AB}$ versus measured input pump power. Black points show experimental results, including $\pm1\sigma$ Bayesian-inferred error bars; blue curves are the model predictions.}
\end{figure*}

Fidelity and EBR results for pump powers from 2.5~mW to 40~mW (the maximum we can achieve with current equipment) are plotted in \cref{fig:exp}. The experimental mean fidelity falls in the interval [0.87,0.93] for all cases, while the EBR increases linearly with power, showing no signs of the turning point expected in theory with sufficient flux (cf. \cref{fig:analysis_curves_surfs}). An accurate estimate of the experimental pair generation flux $x$ requires the pair-production efficiency $\eta_\mathrm{pair}$, defined in our case as the ratio of biphoton pairs produced to the measured input pump flux. This quantity is difficult to obtain due to unknowns in quantum conversion efficiency and the inability to measure facet and waveguide losses independently; our definition $\eta_\mathrm{pair}$ will thus be lower than the intrinsic quantum efficiency, since we consider a larger pump power than what actually enters the waveguide. We can employ a fitting procedure using established coincidence formulas to yield the appropriate values for $\eta_A$, $\eta_B$, and $\eta_\mathrm{pair}$. Combining measured singles and coincidence rates, our procedure yields $\eta_A=(1.2\pm0.2)\times10^{-2}$, $\eta_B=(2.1\pm0.4)\times10^{-4}$, and $\eta_\mathrm{pair}=(2.17\pm0.06)\times10^{-10}$ . Combined with independent measurements of the dark count rates on each detector $d_A = 100$~s$^{-1}$ and $d_B=3500$~s$^{-1}$ and the experimental coincidence window $\tau=1$~ns, we are able to 
compare measurements directly against the curves of \cref{eqn:simple_fidelity,eqn:simple_ebr} predicted by our model, which are likewise included in \cref{fig:exp} (using the mean values of the estimates of $\eta_A$, $\eta_B$, and $\eta_\mathrm{pair}$, which correspond to $y_A=8.33\times10^{-6}$ and $y_B=1.67\times10^{-2}$.



Our fidelities are indeed lower than the theoretical limits for pump powers from 2.5 to 40 mW, over which our model predicts $\mathcal{F}>0.95$. This remaining gap can be attributed to technical limitations from the manual alignment procedures, such as imperfect birefringence compensation and temporal drift in the polarization state through the deployed fiber---nonidealities which are intentionally omitted from the EFA model that focuses on the fundamental contributions of accidental coincidences.
The above considerations notwithstanding, our experimental findings appear consistent with the entanglement distribution model used in our network design simulations. The linear EBR curve suggests that the current QLAN is operating far below peak performance (theory predicts $R_{\max}=1.58\times10^3$~ebits/s for the system noise parameters), and the limited range of fluxes available at present prevents us from demonstrating high-EBR solutions found from the GA procedure. Moving forward, then, it would be valuable to pursue either significantly higher pump powers or alternative hyperentangled biphoton sources based on type-0 phase matching~\cite{Herbauts2013, Vergyris2017}, for which the quantum efficiency should increase by more than 100-fold compared to our type-II source. The model developed and analyzed in this paper thus offers optimism for the near-term realization of flex-grid quantum networks with vastly higher entanglement distribution rates than shown so far; all that is required is an increase in photon flux.

\section{Conclusion}\label{sec:conclusions}
In this paper, we have derived upper bounds on fidelity and EBR for one-to-one links in an optical entanglement-distribution network. These bounds are dependent on the noise parameters of the network users in the form of a dimensionless quantity $y_n$ defined as the ratio of background count probability to system efficiency. Entanglement between users is only possible under certain conditions, namely that their noise parameters $y_1$ and $y_2$ satisfy \cref{eqn:boundary}. 
Using this model, example networks were proposed to test flux allocation optimization using a GA. The number of available frequency channels and the distribution of biphoton flux across those channels have a large influence on the ease of finding a bandwidth allocation that attains near-optimal performance for all links. 
Specifically, noisy links tend to bias the algorithm towards small link fluxes, whereas low-noise links favor higher fluxes. This counteracting pull is absent in networks with links of similar noise parameter pairings. Accordingly, and perhaps unsurprisingly, networks with highly heterogeneous nodes---as evidenced by wide variability in the noise parameters $y_n$---present greater challenges for optimization in a flex-grid quantum network. Irrespective of these challenges, however, by outlining fundamental bounds our model allows us to quantify the closeness between an observed network state and the theoretical ideal, for any bandwidth allocation. This capability should prove invaluable for future network management, enabling a clear distinction between technical and fundamental limitations in reaching a desired configuration.

The model proposed here also has avenues for refinement. The equations and conditions of our dimensionless model are all symmetric to the $y_1=y_2$ line. This may hint at a model parametrization that combines $y_1$ and $y_2$ into a single parameter. Such a parameter would be a good quantitative metric for qualifying a link, as opposed to the relative descriptions of links as ``low'' or ``high'' noise. 
Finally, 
although the current experimental results are consistent with the proposed model, they also indicate operation far below the maximum EBRs anticipated as possible on our QLAN due to limitations on the total amount of biphoton flux we can generate. Consequently, future experiments should attempt to achieve much higher pump powers (or more efficient SPDC) in order to measure EBR beyond the linear, low-flux regime. Good agreement would bode well for subsequent application of GAs in optimizing large-scale entanglement networks.

\section*{Funding}
U.S. Department of Energy, Office of Science, Office of Workforce Development for Teachers and Scientists under the Science Undergraduate Laboratory Internship program; U.S. Department of Energy, Office of Science, Advanced Scientific Computing Research, under the Entanglement Management and Control in Transparent Optical Quantum Networks and Early Career Research programs (Field Work Proposals ERKJ378 and ERKJ353).

\section*{Acknowledgments}
This research was performed in part at Oak Ridge National Laboratory, managed by UT-Battelle, LLC, for the U.S. Department of Energy under contract no. DE-AC05-00OR22725.

\section*{Disclosures}
The authors declare no conflicts of interest.

\section*{Data Availability}
Data available from the authors on request.


\bibliographystyle{jabbrv_aps4-1}
\bibliography{SULIrefs}

\end{document}